\documentclass[11pt]{article}

\usepackage{subcaption}
\usepackage[a4paper,top=2cm,bottom=2cm,left=3cm,right=3cm,marginparwidth=1.75cm]{geometry}
\usepackage[inkscapelatex=false]{svg}
\usepackage{amsmath}
\usepackage{xcolor}
\usepackage{amssymb}
\usepackage{physics}
\usepackage{cleveref}
\usepackage{bm}
\usepackage{empheq}

\graphicspath{{images/}}

\usepackage{amsfonts}
\usepackage{graphicx}
\usepackage{epstopdf}
\usepackage{algorithmic}

\usepackage{amsopn}

\title{Passive magnetoelectric antenna arrays for in vivo monitoring: A theoretical framework \thanks{
This work
was supported by the Air Force Office of Scientific Research under the grant no.  FA9550-23-1-0093 and the AFOSR-MURI program FA9550-25-1-0262.}}

\author{Kalpesh Jaykar\thanks{Department of Aerospace Engineering and Mechanics, University of Minnesota, Minneapolis, MN 
  (jayka001@umn.edu, james@umn.edu).}
\and Prasanth Velvaluri\thanks{Department of Electrical and Computer Engineering, Northeastern University, Boston, MA, 02115, USA
  (p.velvaluri@northeastern.edu).}
\and Nian X. Sun\footnotemark[3]
\and Richard D. James\footnotemark[2]}

\begin{document}

\maketitle

\begin{abstract}

A new brain-computer interface (BCI) technology, deployed through minimally invasive surgery, is changing the way we think about treating severe neurological conditions. The central idea is to place a device called \emph{Stentrode} in the brain’s vasculature, which enables neuromodulation and helps patients regain the ability to communicate. However, in such devices, the battery and electronics are wired and could introduce damage or implant malfunction. In these cases, a Stentrode integrated with magnetoelectric (ME) antennas could be of great interest. ME antennas offer significant advantages over traditional antennas, leveraging acoustic resonance rather than electromagnetic resonance to achieve a size reduction of up to five orders of magnitude. In addition to their compactness and immunity to ground-plane interference, ME antennas could be adopted for use in vascular implants, such as coronary stents, potentially enabling minimally invasive monitoring and communication. Despite these advantages, a single antenna embedded in the implant may be constrained by the limited volume of magnetostrictive material, which could result in low output gain. To address this gain limitation, we propose using antenna arrays designed to produce constructive interference at a designated far-field point, ideally located outside the patient, to enhance signal transmission and receiving capabilities. We develop a mathematical model to represent the antennas and optimize their spatial arrangement and phase synchronization. Simulations based on this model demonstrate promising high-gain performance at the prescribed far-field location through phase manipulation.
\end{abstract}

\section{Introduction}
Smart implants are revolutionizing the medical industry, ranging from brain-computer interfaces (BCIs) to vascular implants. BCIs enable patients with severe disabilities to regain movement, such as controlling an exoskeleton \cite{Vilela2020}, or to communicate via a computer \cite{Vansteensel2016}. Vascular implants, typically used to treat conditions such as strokes and aneurysms, can record vital information like blood pressure \cite{park2019biodegradable_smart_stent} and flow \cite{oyunbaatar2023implantable}, providing critical feedback on implant functionality. Telemetry is an essential component of such implant hardware, enabling communication and/or powering of deeply implanted devices \cite{Malik2021, Nair2023}.

Conventional implant telemetry relies on coils that operate through mutual induction, transmitting and receiving power and data via external coils \cite{Malik2021,Mohammadi_2013}. However, telemetry for vascular implants presents unique challenges due to the minimal available space. Various approaches have been explored, including stents that function entirely as coils \cite{Mohammadi_2013,Liu2019Stent,Shah2023} and stents equipped with dedicated electronics that serve as coils \cite{Zain2025}. These technologies, however, are often too bulky for use in stents designed for brain vasculature to treat aneurysms and strokes, where space is minimal and the device must withstand the crimping process required for minimally invasive delivery via catheters. 

Recently, magnetoelectric (ME) antennas—based on the combination of magnetostrictive and piezoelectric materials—have gained attention due to their compact size and enhanced functionality \cite{Nan2017,chen2022wireless}. Their size reduction stems from their ability to operate at acoustic resonance rather than electromagnetic resonance, allowing for a reduction in size by over five orders of magnitude compared to traditional antennas. Moreover, ME antennas are immune to ground-plane interference, unlike conventional antennas mounted in-plane on conductive substrates \cite{Luo2024}, making them suitable for implants delivered through minimally invasive methods. Two configurations of ME antennas are currently available: one partially released from the substrate \cite{Nan2017}, and another mounted directly on it \cite{Liang2023}. While both have advantages, the latter is more suitable for implants due to its robust construction and superior power-handling capability. These ME antennas can also function as energy harvesters, with an impressive figure of merit of 3721 \cite{zaeimbashi2021ultracompact}.

One limitation in implementing ME antennas in implantable devices is the low gain achieved by a single unit, primarily due to the limited volume of magnetostrictive material \cite{Nan2017}. Vascular devices are typically positioned deep within the brain and must generate sufficient gain to transmit signals to external antennas. To address this challenge, antenna array structures have been explored \cite{Dong2022, Luo2024Magnetoelectric}. At very low frequencies (VLF), the simultaneous excitation of ME antennas enables proportional gain enhancement, as phase shifts due to spatial separation are practically negligible \cite{Dong2022}. While adding more antennas increases the total volume of the magnetostrictive material, the corresponding gain improvement becomes limited at higher frequencies \cite{Luo2024Magnetoelectric}. Moreover, antenna arrays increase the overall device size, which compromises the compactness essential for implantable applications. This is particularly critical in medical applications, where stringent size constraints often prohibit the use of antenna array structures. 

To overcome these limitations without significantly increasing device size, beamforming techniques offer a promising alternative. Beamforming is a well-established method in antenna engineering, where the phase of individual antenna elements is manipulated to direct the radiation pattern toward a desired location \cite{balanis_antenna_1992, Hansen2009}. When applied to ME antenna arrays, phase control can concentrate the radiated energy at a specific far-field point, enhancing signal strength while maintaining compact form factor. In conventional phased arrays, phase control is typically achieved through electronic circuitry \cite{balanis_antenna_1992,mailloux_phased_2005}. However, in implantable applications, electronic phase manipulation may be impractical due to space constraints. It is conceivable that logic gates integrated within or near ME antennas could offer a compact alternative for phase control, though this possibility is not explored in the current work. Nonetheless, the concept introduces a pathway for efficient telemetry in deep-tissue medical implants, where targeted radiation patterns are essential.

A promising approach for implementing implants designed for discrete brain-computer interfaces (BCIs) involves using individual antennas operated with a phase difference\cite{jaykar_macroscopic_2026}. This configuration allows the antennas to generate a point of constructive interference outside the body during operation, resulting in a combined gain greater than that of individual antennas. These antennas can be mounted on the struts of a stent, which can be conceptualized as origami panels \cite{Velvaluri2021,Liu2024}, enabling safe crimping for minimally invasive surgery. Once implanted alongside the stent, the ME antennas may remain inactive until stimulated by an external horn antenna. Upon receiving radiation at their acoustic frequency, the ME antenna arrays respond by emitting their own signal. By measuring the intensity at a designated far-field point, it becomes possible to communicate with the implant and retrieve vital feedback, such as implant positioning or in-vivo data.

The remainder of this paper is organized as follows. \Cref{sec:Background} provides an overview of the background and working principles of magnetoelectric (ME) antennas, highlighting their unique operational characteristics and suitability for implantable applications. \Cref{sec:MathematicalModeling} presents the mathematical modeling framework used to describe the intensity or gain of antenna arrays, including ME antennas. \Cref{sec:PhaseManipulation} explores phase manipulation techniques aimed at achieving constructive interference and enhanced gain at a designated far-field point, introducing a parameterized approach to phase control. \Cref{sec:Results} presents simulation results for proposed optimal antenna placements on a stent structure, considering the relative positioning of the receiver and the desired deployment configuration. Finally, \Cref{sec:FutureScope} outlines potential directions for future research, including practical considerations for implementing ME antennas on stents, such as the possible need for additional electronics to support phase manipulation.

\section{Background and working principle of ME antennas}\label{sec:Background}
Magnetoelectric (ME) materials are composites made by combining piezoelectric and magnetostrictive materials \cite{ryu2001magnetostrictive_layer,dong2005pushpull,greve2010giant,nair2013leadfree}. Piezoelectric materials deform when subjected to an electrical stimulus, a property that arises from asymmetry in certain crystal structures \cite{Priya2017}. Magnetostrictive materials, on the other hand, deform in response to an applied magnetic field, primarily due to the rotation of magnetic domains in alignment with or against the direction of the field \cite{Hristoforou2007}. When these materials are combined, the resulting ME composite exhibits a change in electrical polarization upon the application of a magnetic field—this is known as the direct ME effect. Additionally, an indirect ME effect can occur, where the magnetization in the magnetostrictive layer is modified by an applied electrical field \cite{Fiebig_2005}. 

These ME materials are of particular interest in the design of novel medical implants, including magnetic field sensors \cite{Fiebig_2005, Luo2024Magnetoelectric}, wireless power transfer systems, and miniature electrical simulators. ME antennas are especially attractive for wireless data transfer \cite{Nan2017,Luo2024,zaeimbashi2021ultracompact,Liang2023}; their compact form factor makes them ideal for implants. Unlike conventional antennas that operate based on electromagnetic resonance, ME antennas operate at acoustic resonance, allowing them to be up to six orders of magnitude smaller when operated at the same frequency.

Moreover, ME antennas do not suffer from ground-plane interference, a common issue in conventional antennas mounted in-plane on conductive substrates. In conventional antennas, the image current is out of phase with the primary current, reducing antenna gain. In contrast, ME antennas generate an image current that is in phase with the magnetic current, enhancing antenna gain by 3 dB \cite{Luo2024}.

During reception, ME antennas operate as follows: an external horn antenna emits electromagnetic radiation at the acoustic frequency of the ME antennas. The magnetic layer of the ME antennas responds by deforming in accordance with the magnetic field. This deformation induces strain in the magnetostrictive materials, which is then transferred to the piezoelectric material due to their strain-coupled configuration. The resulting strain in the piezoelectric material generates a proportional radio-frequency (RF) voltage \cite{willcole2022tutorial}.

\section{Mathematical modeling of the antennas}\label{sec:MathematicalModeling}

This section presents a derivation of electromagnetic radiation intensity generated by a finite collection of antennas, each modeled as a classical oscillating charge. In the far-field approximation, the radiation emitted by such an accelerating point charge closely resembles the field produced by a Hertzian dipole. We begin by applying the Li\'enard-Wiechart formalism \cite{griffiths2014} to determine the electric field at a distant observation point $\mathbf{x}$, resulting from a point charge following a trajectory $\mathbf{y}_j (t)$. The expression for a far-field electric field is given by:
\begin{align}
    \mathbf{E}_{out} (\mathbf{x},t) &= \frac{-e}{c^24\pi\epsilon_0} \left( \frac{1}{|\mathbf{x}-\mathbf{y}_j|} \left( \mathbf{I}-\frac{\mathbf{x}-\mathbf{y}_j}{|\mathbf{x}-\mathbf{y}_j|}\otimes\frac{\mathbf{x}-\mathbf{y}_j}{|\mathbf{x}-\mathbf{y}_j|} \right)\Ddot{\mathbf{y}}_j \right)_{t_r} ,
\end{align}
where $e$ is charge, $\epsilon_0$ is permittivity of vaccum, $c$ is speed of light in free space and $(\cdot)_{t_r}$ indicates evaluation at a retarded time $t_r = t-|\mathbf{x}-\mathbf{y}_j|/c$ (see \cref{Appendix:ElectricField}). The Coulomb field is neglected under the far-field assumption. Treating the charge classically, the acceleration of the charge is determined using Newton's second law: $m \Ddot{\mathbf{y}} = \mathbf{F}_{\mathbf{y}_j} e^{-i\omega t}$, where $m$ is the mass of the charge and $\mathbf{F}_{\mathbf{y}_j}$ is the applied force. Grouping constants into a single term $k_{el}$, the electric field simplifies to: 

\begin{align*}
    \mathbf{E}_{out} (\mathbf{x},t) &= k_{el} \frac{1}{|\mathbf{x}-\mathbf{y}_j|} \left( \mathbf{I}-\frac{\mathbf{x}-\mathbf{y}_j}{|\mathbf{x}-\mathbf{y}_j|}\otimes\frac{\mathbf{x}-\mathbf{y}_j}{|\mathbf{x}-\mathbf{y}_j|} \right) \mathbf{F}_{\mathbf{y}_j} e^{-i\omega (t-|\mathbf{x}-\mathbf{y}_j|/c)}\  .
\end{align*}

We now extend this model to a system of multiple antennas, each represented by an oscillating charge at $\mathbf{y}_j(t)$ and driven by an external force $\mathbf{F}_{\mathbf{y}_j}$. Assuming all forces share a common direction and differ only by a phase factor, we set $\mathbf{F}_{\mathbf{y}_j} = \mathbf{F}_0 e^{i \phi_j}$. The total electric field is then obtained by superposing the contribution from each source:
\begin{align*}
    \mathbf{E}_{out} (\mathbf{x},t) &= k_{el} e^{-i \omega t} \sum_j \frac{1}{|\mathbf{x}-\mathbf{y}_j|} \left( \mathbf{I}-\frac{\mathbf{x}-\mathbf{y}_j}{|\mathbf{x}-\mathbf{y}_j|}\otimes\frac{\mathbf{x}-\mathbf{y}_j}{|\mathbf{x}-\mathbf{y}_j|} \right) \mathbf{F}_0 e^{i\phi_j} e^{i(\omega/c)|\mathbf{x}-\mathbf{y}_j|}.
\end{align*}
Under the far-field approximation \cite{Friesecke2016}, where the observation point is much farther than the separation between antennas, we approximate the field using the mean antenna position $\mathbf{y}_s$:
\begin{align*}
    \mathbf{E}_{out} (\mathbf{x},t) &\approx k_{el} e^{-i \omega t} \frac{1}{|\mathbf{x}-\mathbf{y}_s|} \left( \mathbf{I}-\frac{\mathbf{x}-\mathbf{y}_s}{|\mathbf{x}-\mathbf{y}_s|}\otimes\frac{\mathbf{x}-\mathbf{y}_s}{|\mathbf{x}-\mathbf{y}_s|} \right) \mathbf{F}_0 \sum_j e^{i\phi_j} e^{i(\omega/c)|\mathbf{x}-\mathbf{y}_j|},
\end{align*}
Defining the unit vector $\mathbf{n}_s=\frac{\mathbf{x}-\mathbf{y}_s}{|\mathbf{x}-\mathbf{y}_s|}$, the expression simplifies to:
\begin{align} \label{eq:finalelectricfield}
    \mathbf{E}_{out} (\mathbf{x},t) &= k_{el} e^{-i \omega t} \frac{1}{|\mathbf{x}-\mathbf{y}_s|} \left( \mathbf{I}-\mathbf{n}_s\otimes\mathbf{n}_s \right) \mathbf{F}_0 \sum_j  e^{i\phi_j} e^{i(\omega/c)|\mathbf{x}-\mathbf{y}_j|} .
\end{align}

This represents the total electric field generated by the antenna array, where $\mathbf{y}_j$ denotes the position of antenna $j$, $\mathbf{F}_0$ indicates the direction of oscillation for an omnidirectional antenna. The average antenna position is $\mathbf{y}_s$, and $\mathbf{n}_s$ points from the center of arrays to the observation point.

In practical measurement, receivers record the scalar field intensity, defined as the time-averaged magnitude of the Poynting vector, which quantifies electromagnetic energy flux: $$\mathbf{S}(\mathbf{x},t) =\frac{1}{\mu_0} \Re \mathbf{E}_{out} (\mathbf{x},t) \times \Re \mathbf{B}_{out} (\mathbf{x},t),$$ where,  $\mu_0$ is the vacuum permeability, and $\mathbf{B}_{out}$ is the magnetic field. Using the relation $c^2 =\frac{1}{\mu_0 \epsilon_0}$ and the fact that $\mathbf{B}_{out} (\mathbf{x},t) = \frac{\mathbf{n}_s}{c} \cross \mathbf{E}_{out} (\mathbf{x},t)$, we obtain:
\begin{align*}
    \mathbf{S}(\mathbf{x},t) &= \frac{1}{\mu_0} \Re \mathbf{E}_{out} (\mathbf{x},t) \times \Re \mathbf{B}_{out} (\mathbf{x},t)\\
    &= \frac{1}{\mu_0} \left[ \Re  \mathbf{E}_{out}(\mathbf{x},t) \cdot \Re \mathbf{E}_{out}(\mathbf{x},t) \right] \frac{\mathbf{n}_s}{c}\ .
\end{align*}
This result follows from the orthogonality of $\mathbf{E}_{out}(\mathbf{x},t)$ and $\mathbf{n}_s$. The scalar field is then:
\begin{align*}
    I(\mathbf{x}) &= \lim_{T\to\infty} \frac{1}{T} \int_{0}^{T} \abs{\frac{1}{\mu_0} \left[ \Re  \mathbf{E}_{out}(\mathbf{x},t) \cdot \Re \mathbf{E}_{out}(\mathbf{x},t) \right] \frac{\mathbf{n}_s}{c}} dt\\
    &= \frac{1}{\mu_0 c} \lim_{T\to\infty} \frac{1}{T} \int_{0}^{T} \abs{ \Re  \mathbf{E}_{out}(\mathbf{x},t) \cdot \Re \mathbf{E}_{out}(\mathbf{x},t) } dt\  .\\
\end{align*}

\begin{figure}[tbp]
  \centering
  \begin{subfigure}[b]{0.49\textwidth}
    \includegraphics[width=\textwidth]{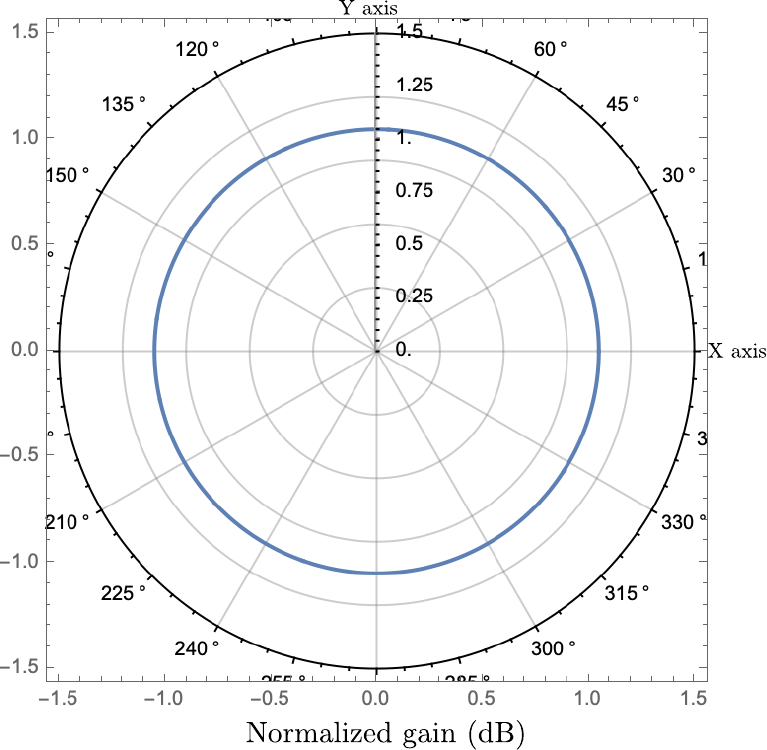}
    \caption{}
    \label{fig:XYPlane}
  \end{subfigure}
  \begin{subfigure}[b]{0.49\textwidth}
    \includegraphics[width=\textwidth]{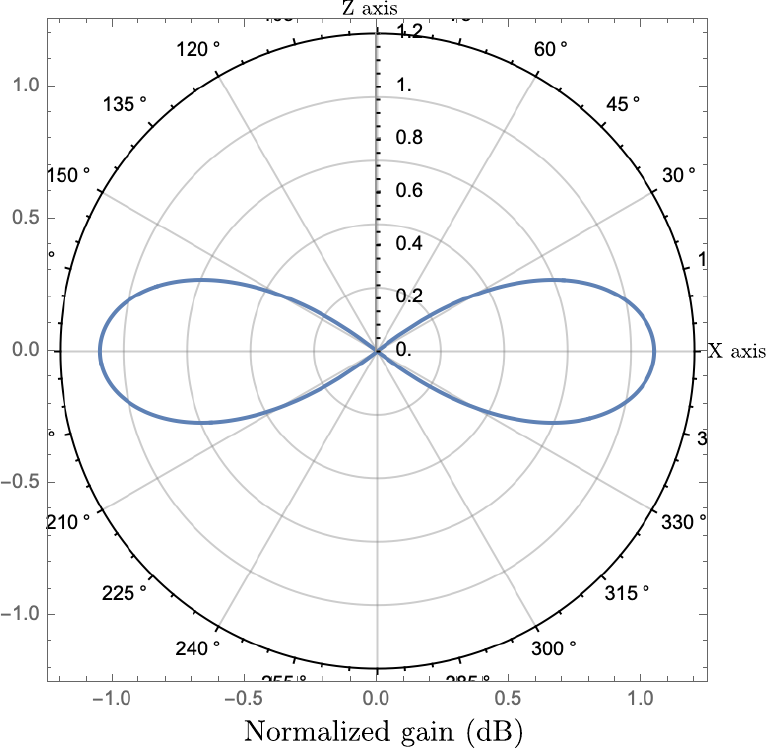}
    \caption{}
    \label{fig:XZPlane}
  \end{subfigure}
  \begin{subfigure}[b]{0.49\textwidth}
    \includegraphics[width=\textwidth]{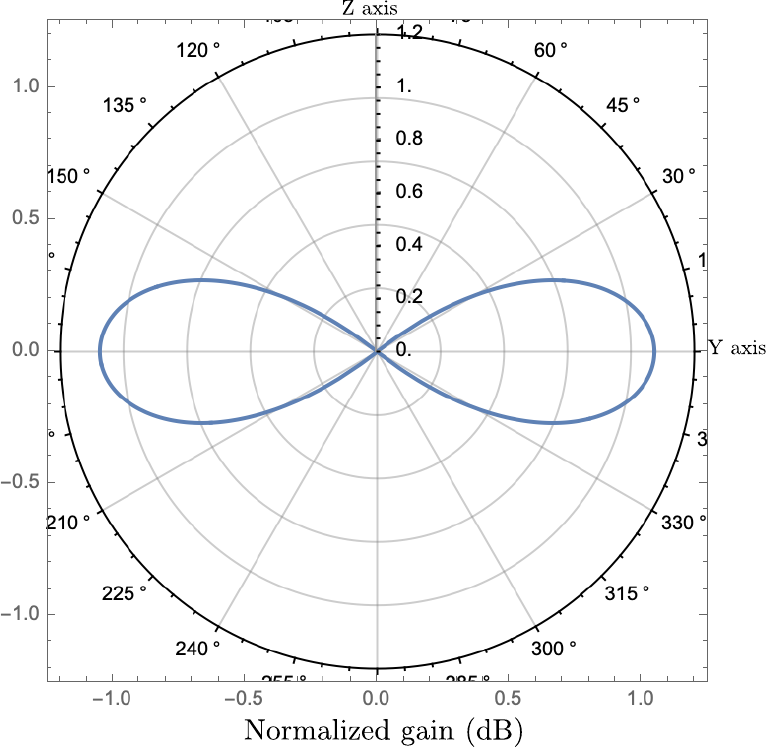}
    \caption{}
    \label{fig:YZPlane}
  \end{subfigure}
  \begin{subfigure}[b]{0.49\textwidth}
    \includegraphics[width=\textwidth]{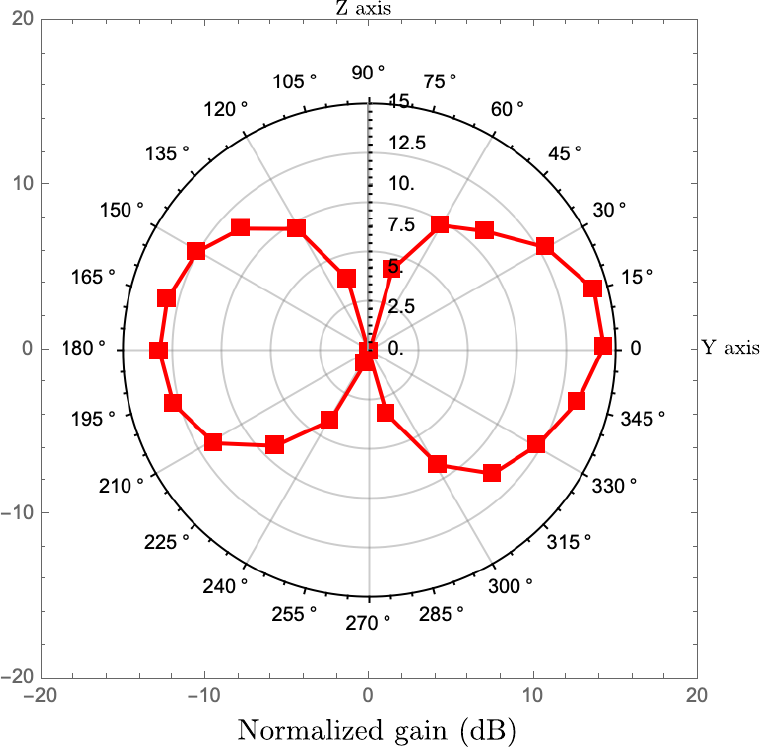}
    \caption{}
    \label{fig:GainPlotExp}
  \end{subfigure}
  \caption{Analytical gain plots of the ME antenna in three orthogonal planes (\protect\subref{fig:XYPlane})XY, (\protect\subref{fig:XZPlane})XZ, and (\protect\subref{fig:YZPlane})YZ, generated using Mathematica. Subplot (\protect\subref{fig:GainPlotExp}) presents the experimental gain measurement \cite{Nan2017} in the YZ plane. The analytical and experimental results show qualitative agreement, supporting the validity of the proposed mathematical model.} \label{fig:GainPlot}
\vspace{-8mm}
\end{figure}

Letting $\mathbf{E}_{out} (\mathbf{x},t) = (\mathbf{R}(\mathbf{x}) + i \mathbf{P}(\mathbf{x})) e^{-i \omega t}$, where $\mathbf{R}(\mathbf{x})$ and $\mathbf{P}(\mathbf{x})$ are the real and imaginary components of the complex electric field, respectively, we compute:
\begin{align*}
    I(\mathbf{x}) &= c\epsilon_0\lim_{T\to\infty} \frac{1}{T} \int_{0}^{T} [|\mathbf{R}(\mathbf{x})|^2 \cos^2(\omega t) + |\mathbf{P}(\mathbf{x})| \sin^2(\omega t) + 2 \mathbf{R}(\mathbf{x})\cdot \mathbf{P}(\mathbf{x}) \sin(\omega t) \cos(\omega t)] dt\\
    &= c\epsilon_0 \left[ \frac{1}{2} |\mathbf{R}(\mathbf{x})|^2 + \frac{1}{2} |\mathbf{P}(\mathbf{x})|^2 \right]\\
    &= \frac{c\epsilon_0}{2} |\mathbf{E}_{out}(\mathbf{x})|^2.
\end{align*}

Substituting from \Cref{eq:finalelectricfield}, the intensity at point $\mathbf{x}$ becomes:
\begin{align}\label{eq:finalintensity}
    I(\mathbf{x}) &= \frac{ c\epsilon_0 k_{el}^2}{2 |\mathbf{x}-\mathbf{y}_s|^2} \left| \left( \mathbf{I}-\mathbf{n}_s\otimes\mathbf{n}_s \right) \mathbf{F}_0 \right|^2 \left| \sum_j  e^{i\phi_j} e^{i(\omega/c)|\mathbf{x}-\mathbf{y}_j|}\right|^2 \ .
\end{align}

\section{Phase manipulation for high-gain output}\label{sec:PhaseManipulation}

Having established the mathematical formulation of the electromagnetic field generated by an array of antennas, we now explore how phase manipulation can be used to optimize the radiation intensity at a desired far-field location. This section focuses on strategies for achieving constructive interference through controlled phase offsets.

\begin{figure}[htbp]
  \centering
  \begin{subfigure}[b]{0.49\textwidth}
    \includegraphics[width=\textwidth]{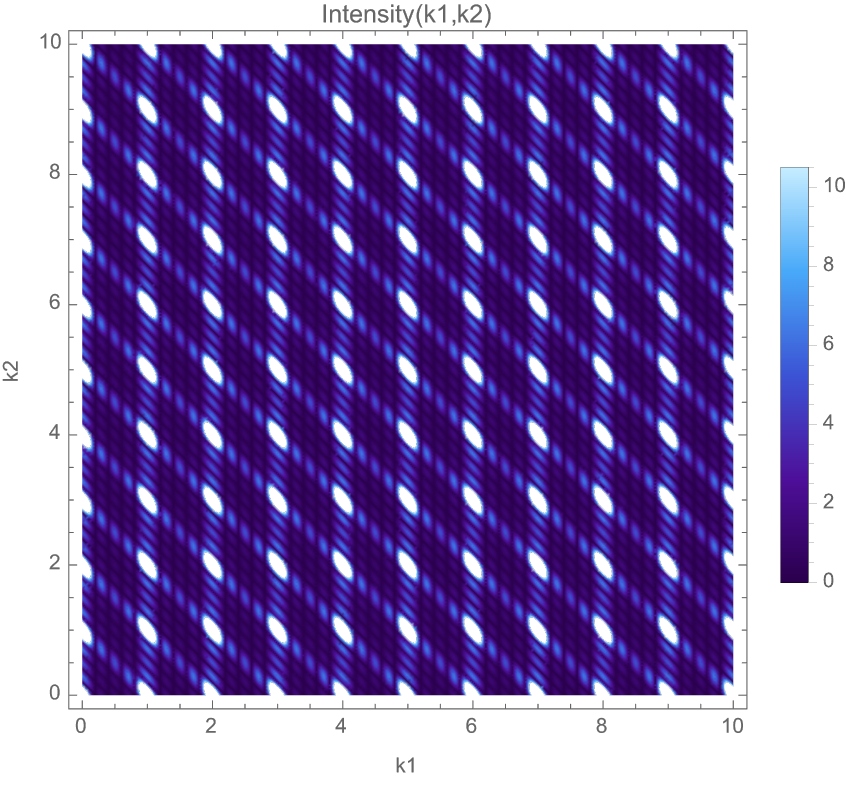}
    \caption{}
    \label{fig:intensityplots1}
  \end{subfigure}
  \begin{subfigure}[b]{0.49\textwidth}
    \includegraphics[width=\textwidth]{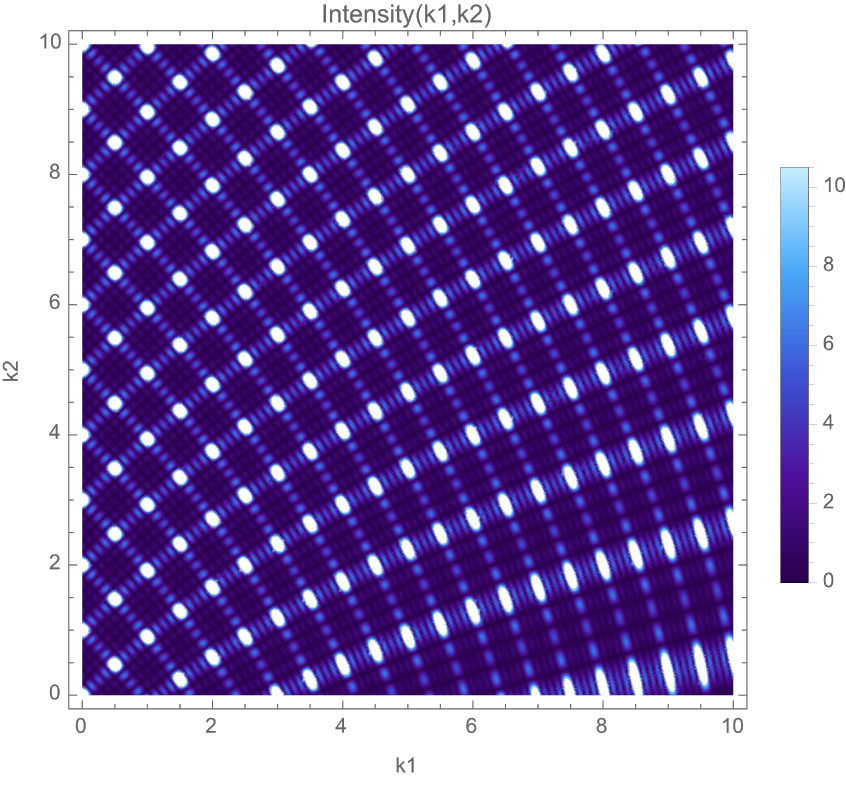}
    \caption{}
    \label{fig:intensityplots2}
  \end{subfigure}
  \begin{subfigure}[b]{0.98\textwidth}
    \includegraphics[width=\textwidth]{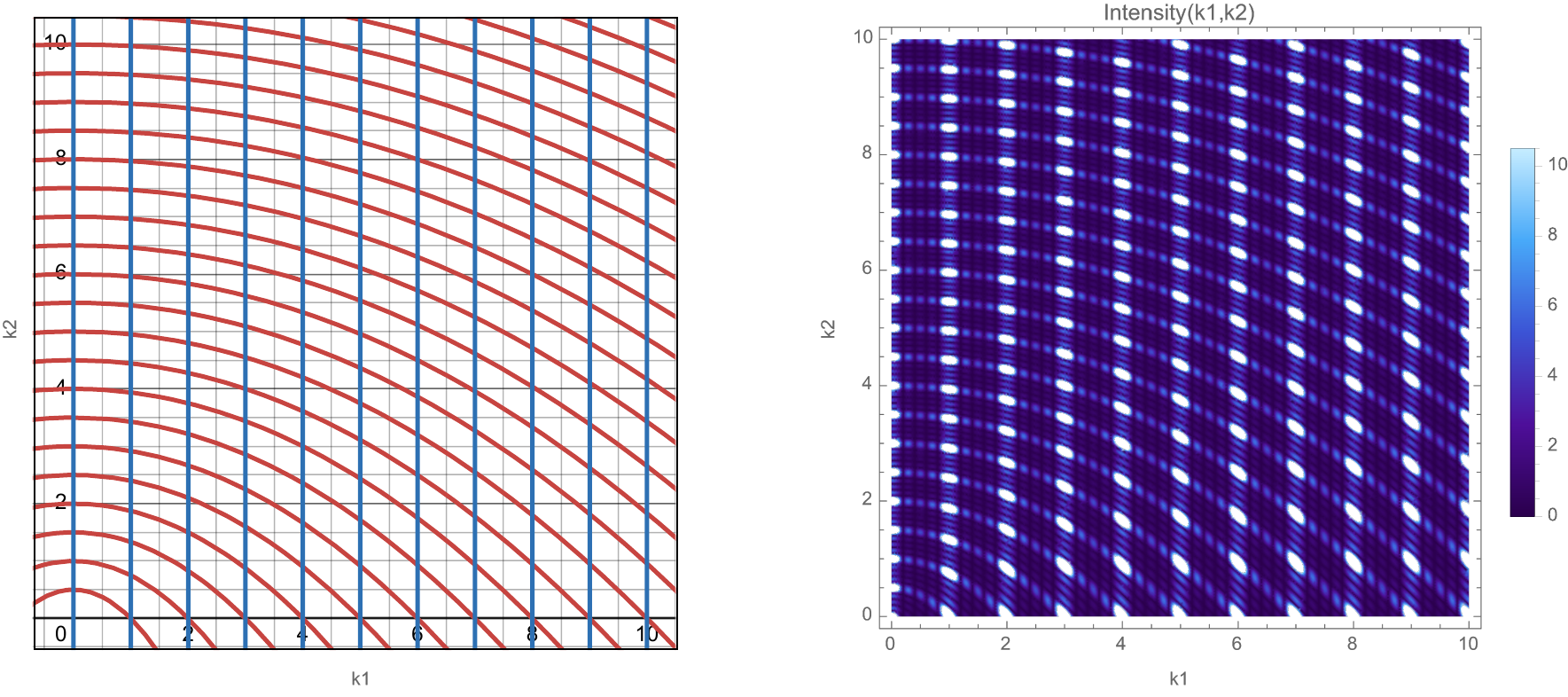}
    \caption{}
    \label{fig:intensityplots3comparison}
  \end{subfigure}
  \caption{Intensity plots as a function of parameters $k_1$ and $k_2$ for 
(\protect\subref{fig:intensityplots1})  $\varphi_{lm} = 2 \pi \left((m+l) k_1  + l k_2\right)$ and (\protect\subref{fig:intensityplots2}) $\varphi_{lm} = 2 \pi \left((m+l) k_1  + (l-m)\sqrt{k_1^2+(k_2+4)^2}\right)$. (\protect\subref{fig:intensityplots3comparison}) Left: Analytical plots for two families of parameter lines corresponding to partial constructive interference: $k_2 + \sqrt{k_1^2 + k_2^2} = \mathbb{Z}$ (red) and $k_1 =\mathbb{Z}$ (blue). The intersection of these lines represents the parameter values corresponding to the highest intensity, where constructive interference is strongest. Right: the intensity plots of the electromagnetic field for $\varphi_{lm} = 2 \pi \left(m k_1 + l k_2 + l \sqrt{k_1^2 + k_2^2}\right)$ illustrating high intensity regions aligned with the analytical predictions.} \label{fig:intensityplots}
\vspace{-8mm}
\end{figure}

Let $\mathbf{y}_j$ denotes the mean position of antenna $j$. Given the position of the receiver $\mathbf{x}$ and individual antenna positions $\mathbf{y}_j$, it is possible to adjust the phase $\phi_j$ of each antenna to shape the resulting intensity pattern. The phase delays naturally arise from the propagation term $e^{i (\omega/c) \left| \mathbf{x} - \mathbf{y}_j \right|}$ in the field expression. By selecting $\phi_j = - (\omega /c) |\mathbf{x}-\mathbf{y}_j| $, the radiated fields from all antennas arrive in phase at the receiver, maximizing the observed intensity.

More generally, the same constructive interference condition can be achieved by setting $\phi_j = 2\pi \mathbb{Z} - (\omega /c) |\mathbf{x}-\mathbf{y}_j| $. This allow us to define a parameterize phase function $\phi_j(k_1,k_2) = \varphi_j(k_1,k_2) - (\omega /c) |\mathbf{x}-\mathbf{y}_j|$, where $k_1$ and $k_2$ are tunable parameters. The resulting intensity becomes a function of these parameters, enabling flexible control over the interference pattern. In \Cref{fig:intensityplots}, we illustrate how different choices of $\varphi_j(k_1,k_2)$ influence the intensity distribution.

\begin{figure}[htbp]
  \centering
  \includegraphics[width=0.6\textwidth]{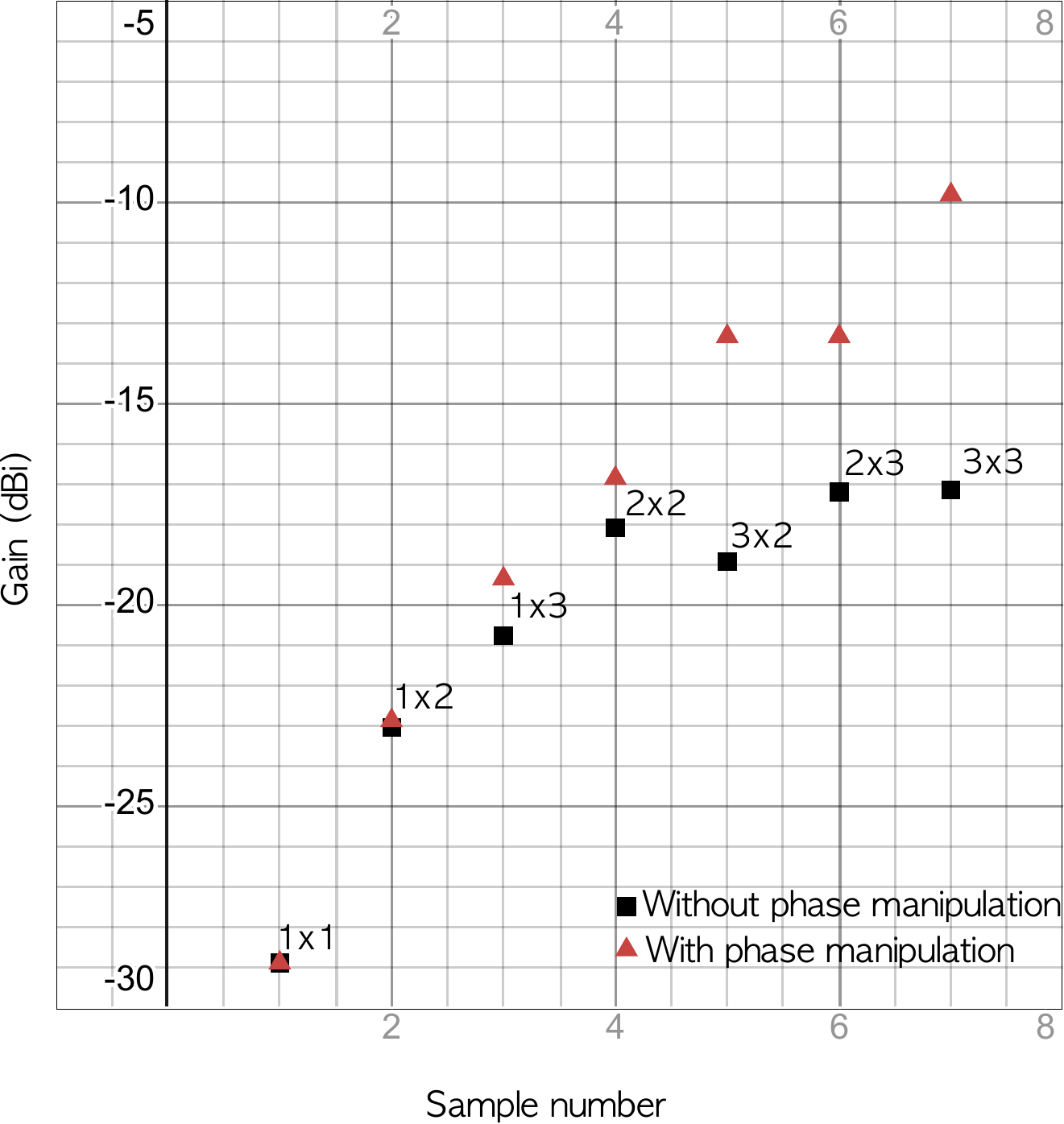}
  \caption{Antenna gain plots for various ME antenna array configurations (adapted from Luo et al. \cite{Luo2024}). The results show that increasing the number of antennas does not lead to proportional gain enhancement. This behavior is attributed to the individual signals from ME antennas being out of phase at the receiver location, resulting in partial or destructive interference.} \label{fig:AntennaArrayData}
\vspace{-8mm}
\end{figure}

To interpret the plots in \cref{fig:intensityplots}, consider a phase function of the form $\varphi_{lm} = l f_l (k_1,k_2) + m f_m (k_1,k_2)$. Three distinct interference regimes emerge:

\begin{itemize}
    \item \textbf{Full constructive interference} occurs when both $f_l(k_1,k_2)$ and $f_m(k_1,k_2)$ are integer multiples of $2\pi$. In this case, all antennas are phase-aligned at the receiver, and the intensity scales quadratically with the number of antennas:
    \begin{align*}
        \left| \sum_{l,m}  e^{i\varphi_{lm}} \right|^2 = (\#l)^2 (\#m)^2 \ .
    \end{align*}

    \item \textbf{Partial constructive interference} arises when only one of the functions, say $f_l(k_1,k_2)$, satisfies the $2\pi \mathbb{Z}$ condition. This leads to subsets of antennas being in phase, but with relative phase different between the subsets:
    \begin{align*}
        \left| \sum_{l,m}  e^{i\varphi_{lm}} \right|^2 = (\#l)^2 \left| \sum_{m}  e^{i m f_m (k_1,k_2)} \right|^2.
    \end{align*}
    The second term diminishes as $m$ increases, resulting in reduced intensity.

    \item \textbf{Destructive interference} dominates when neither $f_l(k_1,k_2)$ nor $f_m(k_1,k_2)$ are integer multiples of $2\pi$. In this case, the summation terms decay rapidly, and the antennas are out of phase at the receiver (see also \Cref{fig:AntennaArrayData}).
    \end{itemize}

These results demonstrate how phase manipulation, guided by parameterized functions, can be used to engineer constructive interference patterns.

\section{Results and discussions}\label{sec:Results}

This section presents the central idea of the paper: using ME antennas mounted directly on a substrate in an array configuration, to achieve high-gain wireless transmission for smart implants that support minimally invasive implantation. ME antennas are compact and well-suited for implantable applications, but a single unit does not generate sufficient gain for reliable telemetry. A natural solution is to use multiple antennas arranged in an array, all operating in phase, to boost the signal strength.

At lower frequencies, this approach works well—arrays of ME antennas can achieve proportional gain enhancement, as shown in \cite{Dong2022}. However, at higher frequencies, the gain does not scale as expected (see \Cref{fig:AntennaArrayData}). The reason is that the phase differences introduced by the spatial separation of antennas become significant due to the shorter wavelength. These phase shifts arise naturally from the physics of wave propagation. As a result, the signals from individual antennas arrive out of phase at the receiver, leading to partial or destructive interference.

To overcome this limitation, we propose a phase manipulation strategy, as described in \Cref{sec:PhaseManipulation}, that compensates for the phase shifts caused by spatial separation. By carefully adjusting the relative phase of each antenna, we can ensure that the radiated signals arrive in phase at a designated far-field receiver location, restoring constructive interference and enabling high-gain transmission.

\begin{figure}[htbp]
  \centering
    \includegraphics[height=0.9\textheight]{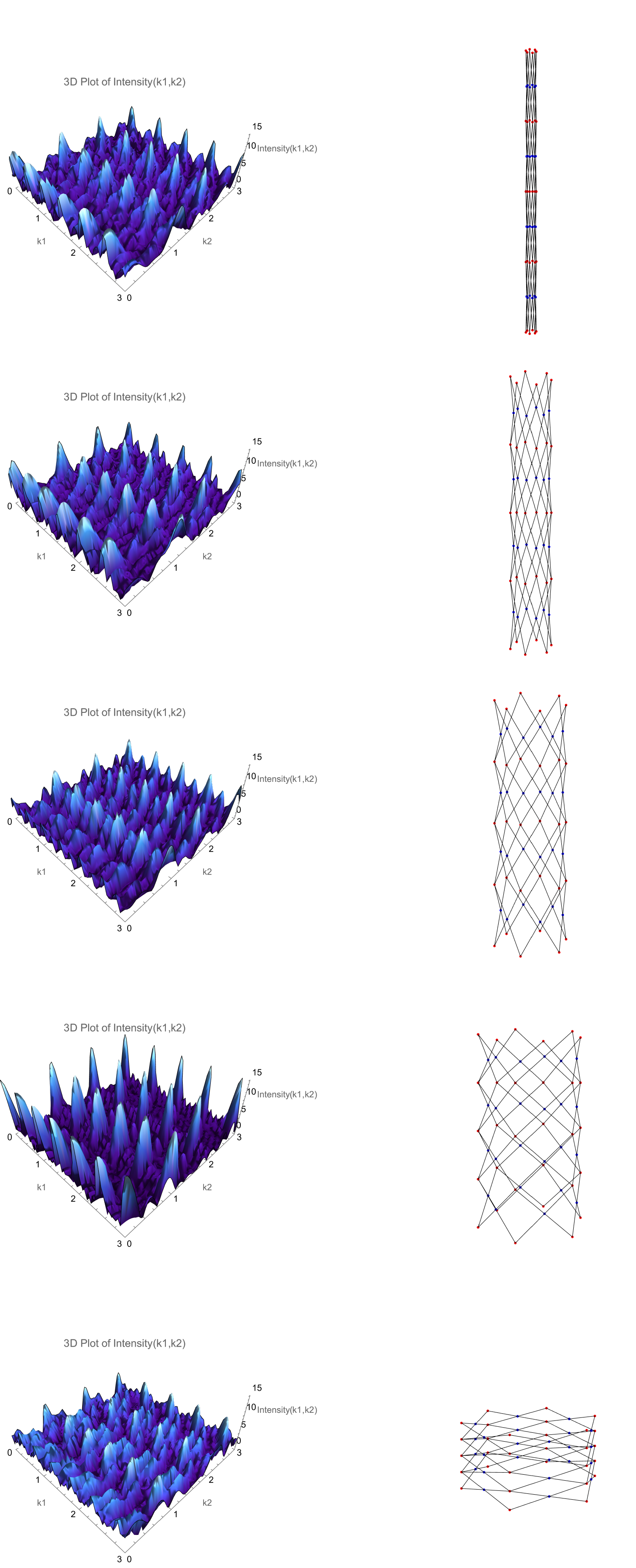}
  \caption{Simulated intensity plots showing signal strength as a function of phase parameters $k_1$ and $k_2$. Peaks indicate configurations where ME antennas, mounted on the stent, produce constructive interference at the receiver. These values guide phase selection prior to implantation for deployment verification. } \label{fig:intensityplotsMEAntennas}
\vspace{-8mm}
\end{figure}

The relative phases of the ME antennas are defined before implantation using the parameterized phase functions introduced earlier. These phases are determined by selecting values of $k_1$ and $k_2$ that yield a peak in intensity for the desired stent configuration. The intensity plots in \Cref{fig:intensityplotsMEAntennas} represent the most critical result of this study. They demonstrate how phase manipulation enables constructive interference at the receiver despite spatial separation. Once a suitable pair ${k_1, k_2}$ is chosen, the corresponding phase offsets are assigned to the antennas to ensure high-gain transmission.

\begin{figure}[htbp]
  \centering
  \includegraphics[width=\textwidth]{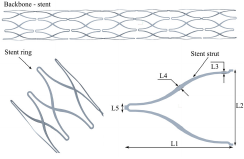}
  \caption{Possible mounting locations for ME antennas on the stent structure (adapted from Velvaluri et al. \cite{Velvaluri2021ThinFilm}). Among the identified regions, the location labeled “L4” offers the greatest width, providing the largest available surface area for antenna placement. For the purpose of analysis, ME antennas are assumed to be positioned circumferentially along the L4 region and at multiple longitudinal positions. This configuration is selected based on its suitability for accommodating ME antennas while maintaining structural integrity during crimping.} \label{fig:MEAntennaPosition}
\end{figure}

Integrating ME antennas into stents introduces additional mechanical constraints. During minimally invasive surgery, the stent must be compressed to a smaller diameter and then expanded to its final shape inside the body. This process subjects the stent—and any mounted components—to significant crimping forces. To ensure mechanical robustness and sufficient surface area for mounting, we adopt the antenna placement strategy proposed in \cite{Velvaluri2021ThinFilm}. Specifically, we focus on the L4 region of the stent, which offers the greatest width and structural support for antenna integration (see \Cref{fig:MEAntennaPosition}). In addition to its greatest width, finite element simulations show that the stress generated in this region is minimal during the crimping process, helping ensure that the antennas remain undamaged during loading and unloading from the catheter.

ME antennas may remain passive at all times. However, after deployment, they can be activated by an external horn antenna  that emits radiation at their acoustic frequency. Upon receiving this radiation, the antennas can function as energy harvesters, potentially powering small electronics mounted on the stent. If sufficient energy is stored using a capacitor, it can then be used to power the ME antennas as transmitters. During transmission, the antennas could transmit data related to implant positioning or key functional parameters. To ensure meaningful detection, the receiver must be appropriately positioned relative to a known anatomical reference point within the patient’s body and therefore relative to the implant, as discussed in the previous section. If the antennas are correctly configured, signals from individual ME antennas will arrive in phase, resulting in strong gain. This would confirm that the ME antennas—and hence the stent—are in their intended configuration. This method provides a non-invasive means of verifying implant positioning, eliminating the need for procedures such as endoscopy. The simulated results are shown in \Cref{fig:intensityplotsMEAntennas}, where stents are represented by black wireframes, and the ME antenna positions are indicated by solid blue and red dots.
In summary, this approach enables three key functionalities:

\begin{itemize}
    \item Deployment verification: By checking for peak intensity at a known receiver location, we can confirm whether the stent has been deployed correctly, as required by the physician.
    \item Post-deployment monitoring: The same setup can be used periodically to monitor the stent’s position and integrity over time.
    \item Long-term functionality: A synchronous antenna system mounted on the stent could enable bidirectional data transmission, allowing the device to act as a discrete implantable unit capable of brain recording and stimulation.
\end{itemize}

Although the results are promising, several practical challenges remain.N

\section{Future works}\label{sec:FutureScope}
While the proposed approach shows promise for using ME antennas in smart vascular implants, several challenges remain before practical deployment can be achieved.

The most immediate challenge is the fabrication of such novel devices and the complexities associated with it. The biocompatibility of the various functional layers in the ME antennas must be thoroughly verified. Additionally, the antennas should be propperly packaged to:
\begin{enumerate}
    \item protect them from foreign body reactions, and
    \item safeguard the body in case the antenna materials are not inherently biocompatible.
\end{enumerate}

Following fabrication, the most critical issue is phase control. In conventional antenna systems, integrated circuits (ICs) are commonly used to manipulate phase electronically. However, ME antennas mounted directly on the substrate face stringent space constraints, making it difficult to incorporate such circuitry. Developing compact, low-power mechanisms for phase control—possibly through material-level or structural innovations—remains an open challenge.

Another limitation lies in the assumptions made during modeling. The mathematical framework used to optimize phase manipulation assumes a homogeneous, charge-free medium. In reality, the human body is highly heterogeneous and electrically active, which can introduce signal attenuation and phase distortion \cite{Abbas2022}. Future work must account for these biological complexities, possibly through in-vivo calibration or adaptive phase tuning.

Additionally, current ME antenna designs offer limited control over their resonance frequency. This restricts flexibility in tuning the system for different diagnostic or communication needs. Exploring tunable ME structures or hybrid designs could help address this constraint.

Finally, the intensity peaks achieved through phase manipulation are highly sensitive to the deployment configuration. Ideally, the system should tolerate small deviations in stent placement and still produce a reasonably strong signal. However, the current approach may result in sharp drops in intensity if the deployment deviates even slightly from the target configuration. Designing more robust phase functions or incorporating tolerance-aware optimization could improve reliability in real-world scenarios.

\noindent{\bf Acknowledgment.} Kalpesh Jaykar and Richard D. James acknowledge the support from the Air Force Office of Scientific Research under grant number FA9550-23-1-0093 and the AFOSR-MURI program FA9550-25-1-0262. Prasanth Velvaluri and Nian Sun acknowledge funding from the Alexander von Humboldt Foundation through the Feodor Lynen Fellowship. 

\bibliography{ApplicationOfMeAntennas}
\bibliographystyle{ieeetr}

\appendix
\section{Electric field produce by an accelerating charge}\label{Appendix:ElectricField}

For completness we outline the Li\'enard-Wiechart method of generating electromagnetic fields in the context of this paper.  This method relies on the fact that electromagnetic radiation takes finite time to travel a given distance. In such cases, the retarded time is defined as $t_r = t-\frac{1}{c}|\mathbf{x}-\mathbf{y}_j|$, where $\mathbf{x}$ is the observation point, $\mathbf{y}_j$ is the observed point subject to the variations of source charge, $c$ is the speed of electromagnetic radiations in free space, $t$ is time and $|\cdot|$ is Euclidean norm. For a moving point-charge, the trajectory $\mathbf{y}_j$  is a function of retarded time. Thus, following the trajectory of charge, the implicit function of retarded time is given by $$t_r=t-\frac{1}{c}|\mathbf{x}-\mathbf{y}_j(t_r)|.$$ The scalar and vector potential ($\phi(\mathbf{x},t) $ and $\mathbf{A}(\mathbf{x},t)$) satisfy the non homogeneous wave equation where the sources have given charge and current densities $\rho(\mathbf{x},t)$ and $\mathbf{J}(\mathbf{x},t)$. Using the Lorenz gauge, we get the uncoupled and symmetric non-homogeneous wave equations:

\begin{align*}
    \grad^2 \varphi -\frac{1}{c^2} \frac{\partial^2 \varphi}{\partial t^2}&= -\frac{\rho}{\epsilon_0}\\
    \grad^2 \mathbf{A} -\frac{1}{c^2} \frac{\partial^2 \mathbf{A}}{\partial t^2}&= -\mu_0 \mathbf{J}\\
\end{align*}

This can be solved to obtain the potentials in terms of retarded time. For a single moving charge with charge $e$ whose trajectory is given as a function of time by $\dot{\mathbf{y}_j}(t)$, the charge and current densities are $\rho(\mathbf{x},t)= e \delta^3 \left(\mathbf{x}-\mathbf{y}_j(t)\right)$ and $\mathbf{J}(\mathbf{x},t) = e \dot{\mathbf{y}_j}(t_r) \delta^3 \left(\mathbf{x}-\mathbf{y}_j(t)\right)$. Here, $\delta^3$ is the three-dimensional Dirac delta function. Using $\mathbf{n}_j = (\mathbf{x}-\mathbf{y}_j)/|\mathbf{x}-\mathbf{y}_j| $ and $\bm{\beta}_j(t_r) = \dot{\mathbf{y}_j}(t_r)/c $, we get:
\begin{align*}
    \varphi(\mathbf{x},t)&=\frac{1}{4\pi\epsilon_0} \left(\frac{e}{(1-\mathbf{n}_j\cdot\bm{\beta}_j)|\mathbf{x}-\mathbf{y}_j|} \right)_{t_r}\\
    \mathbf{A}(\mathbf{x},t)&=\frac{\mu_0 c}{4\pi} \left(\frac{e \bm{\beta}_j}{(1-\mathbf{n}_j\cdot\bm{\beta}_j)|\mathbf{x}-\mathbf{y}_j|} \right)_{t_r}\\
\end{align*}
where, $\epsilon_0$ is permittivity of free space and $(\cdot)_{t_r}$ means that the quantities in parentheses should be evaluated at the retarded time $t_r$. The fields can then be easily calculated since $\mathbf{E}_{out} = -\grad{\varphi} - \frac{\partial\mathbf{A}}{\partial t}$ and $\mathbf{B}_{out} = \curl{\mathbf{A}}$. Using $\gamma_j (t) = \frac{1}{\sqrt{1-|\bm{\beta}_j (t)|^2}}$, we get : 

\begin{align*}
    \mathbf{E}_{out} (\mathbf{x},t) = \frac{1}{4\pi\epsilon_0} \left( \frac{e(\mathbf{n}_j-\bm{\beta}_j)}{\gamma_j^2(1-\mathbf{n}_j\cdot\bm{\beta}_j)^3|\mathbf{x}-\mathbf{y}_j|^2} + \frac{e\mathbf{n}_j\times((\mathbf{n}_j-\bm{\beta}_j)\times\dot{\bm{\beta}_j})}{c(1-\mathbf{n}_j\cdot\bm{\beta}_j)^3|\mathbf{x}-\mathbf{y}_j|} \right)_{t_r}
\end{align*}

When the speed of the charge is small compared to the speed of electromagnetic radiations in free space, we can assume $|\bm{\beta}_j| \ll 1$ and hence $\gamma_j \approx 1, \quad 1-\mathbf{n}_j\cdot\bm{\beta}_j \approx 1, \quad \mathbf{n}_j-\bm{\beta}_j \approx \mathbf{n}_j, \quad 1-|\bm{\beta}_j|^2\approx 1 $. This  simplifies the electric field:

\begin{align*}
    \mathbf{E}_{out} (\mathbf{x},t) &= \frac{1}{4\pi\epsilon_0} \left( \frac{e(\mathbf{n}_j-\bm{\beta}_j)}{\gamma_j^2(1-\mathbf{n}_j\cdot\bm{\beta}_j)^3|\mathbf{x}-\mathbf{y}_j|^2} + \frac{e\mathbf{n}_j\times((\mathbf{n}_j-\bm{\beta}_j)\times\dot{\bm{\beta}_j})}{c(1-\mathbf{n}_j\cdot\bm{\beta}_j)^3|\mathbf{x}-\mathbf{y}_j|} \right)_{t_r}\\
    &\approx \frac{1}{4\pi\epsilon_0} \left( \frac{e\mathbf{n}_j}{|\mathbf{x}-\mathbf{y}_j|^2} + \frac{e\mathbf{n}_j\times(\mathbf{n}_j\times\dot{\bm{\beta}_j})}{c|\mathbf{x}-\mathbf{y}_j|} \right)_{t_r}\\
    &= \frac{e}{4\pi\epsilon_0} \left( \frac{\mathbf{n}_j}{|\mathbf{x}-\mathbf{y}_j|^2} + \frac{(\mathbf{n}_j\cdot\dot{\bm{\beta}_j})\mathbf{n}_j-(\mathbf{n}_j\cdot\mathbf{n}_j)\dot{\bm{\beta}_j})}{c|\mathbf{x}-\mathbf{y}_j|} \right)_{t_r}\\
    &= \frac{e}{4\pi\epsilon_0} \left( \frac{\mathbf{n}_j}{|\mathbf{x}-\mathbf{y}_j|^2} - \frac{(\mathbf{I}-\mathbf{n}_j\otimes\mathbf{n}_j)\dot{\bm{\beta}_j}}{c|\mathbf{x}-\mathbf{y}_j|} \right)_{t_r}\\
\end{align*}

The first term can be recognized as the Coulomb electric field due to presence of the charge. The second term is also a well-known field, which is due to acceleration of the charge\cite{Miller1980}. In the far field, the Coulomb field vanishes and  the field due to acceleration dominates. Re-substituting $\mathbf{n}_s = (\mathbf{x}-\mathbf{y}_s)/|\mathbf{x}-\mathbf{y}_s| $ and $\bm{\beta}(t_r) = \dot{\mathbf{y}_s}(t_r)/c $ and considering the far field, we get

\begin{align} \label{eq:accfield}
    \mathbf{E}_{out} (\mathbf{x},t) &= \frac{-e}{c^24\pi\epsilon_0} \left( \frac{1}{|\mathbf{x}-\mathbf{y}_j|} \left( \mathbf{I}-\frac{\mathbf{x}-\mathbf{y}_j}{|\mathbf{x}-\mathbf{y}_j|}\otimes\frac{\mathbf{x}-\mathbf{y}_j}{|\mathbf{x}-\mathbf{y}_j|} \right)\Ddot{\mathbf{y}}_j \right)_{t_r}
\end{align}

\end{document}